\documentclass{gandb}
\usepackage{epsfig}
\usepackage{float}

\let\<=\langle
\let\>=\rangle
\def\vec#1{\mbox{\boldmath{$#1$}}}
\def\div{\mathop{\mathrm{div}}\nolimits}
\begin{document}
\title{IMPROVED ENVELOPE AND EMITTANCE DESCRIPTION}
{IMPROVED ENVELOPE AND EMITTANCE\protect\\
DESCRIPTION OF PARTICLE BEAMS\protect\\
USING THE FOKKER-PLANCK APPROACH}
\author{J\"URGEN STRUCKMEIER}
\authorhead{J.~STRUCKMEIER}
\address{\it Gesellschaft f\"ur Schwerionenforschung (GSI),
Postfach~11~05~52,\protect\\ 64220~Darmstadt, Germany}
\received{(Received 28 October 1993; in final form 5 April 1994)}
\abstract{Beam dynamics calculations based on the collisionless
Vlasov equation omit collisional correlations such as intra-beam scattering.
For a short-correlated process dominated by small momentum changes,
a Fokker-Planck equation can describe their effect on the one-particle distribution.
From this equation we derive equations of motion for the beam envelopes
and for the rms-emittances.
Prescribed friction against an external bath is distinguished from
momentum- and energy-conserving self-collisions.
For the latter, an affine mean-flow approximation and a scalar relaxation
coefficient give simple equations for temperature balancing within the beam.
The longitudinal diffusion and temperature are expressed in terms of the
physical random momentum rather than the signed effective longitudinal mass.
The resulting growth-rate formulas apply to this reduced relaxation model.
The original numerical examples are retained for comparison, with the
limitations of the storage-ring and numerical-noise applications made explicit.}
\keywords{envelope equation, Fokker-Planck equation,
storage ring, intra-beam scattering\newline~\newline
\hspace*{-22mm}Revised version of: Particle Accelerators~{\bf 45}, 229 (1994)}
\maketitle
%\begingroup\footnotesize
%\noindent Corrected working version, 16 October 2026.
%The historical publication information and figures are retained.
%The numerical examples have not been rerun with the corrected model.
%\par\endgroup
\section{INTRODUCTION}
Analytical approaches in the optics of charged particle beams
including self-forces of the beam usually
assume that Liouville's theorem applies for the phase
space distribution function $f$ in the $6$-dimensional `$\mu$~space'.
According to this theorem, the volume of an element of phase space
populated by particles remains constant in time.
From Liouville's theorem we can derive straightforwardly the
equation of motion for $f$, commonly referred to as
the Vlasov or the `collisionless' Boltzmann equation in the general case.
Although direct integration of the
Vlasov equation is generally difficult, it may serve as the
starting-point for analytical investigations under certain conditions.
As an example, we quote the so-called
`Kapchinskij-Vladimirskij' model\cite{kapvla}
for the distribution function $f$, which simplifies the Vlasov
equation to yield a closed set of differential equations
for the envelopes of an unbunched beam.

Liouville's theorem concerns Hamiltonian phase-space flow.
It applies to the full many-particle system even when the particles interact.
The collisionless Vlasov equation is a mean-field equation for the
one-particle distribution and does not include collisional correlations.
Collisionless phase mixing and nonlinear mean fields can nevertheless
increase rms-emittances without violating fine-grained Liouville conservation.
The present extension concerns a reduced kinetic description of
collisional relaxation, rather than a failure of the many-particle theorem.

Recently, Bohn\cite{bohn} made an attempt to extend
the self-consistent Vlasov-Poisson treatment of beam optics.
He included the Fokker-Planck equation in his analysis to describe
the dilution of the phase space due to turbulent heating for a
mismatched `sheet' beam.

The Fokker-Planck equation is generally more complicated than the Vlasov equation.
Moment equations provide a useful alternative to its direct integration.
The complexity of the problem of directly integrating the
Vlasov equation led to attempts to simplify the
analytical description of beam optics.
A largely accepted method has been presented by
Lapostolle\cite{lapostolle} and Sacherer\cite{sacherer}
describing the beam optics in terms of `root-mean-square' (rms)
beam moments and their respective equations of motion.
In this paper, we apply this idea to derive
`moment equations' to the Fokker-Planck equation.

We thus obtain additional moment terms that --- at least in principle ---
allow us to describe short-correlated drift and diffusion processes within the beam
not yet covered by the collisionless mean-field approach.
The remaining task is to determine the Fokker-Planck
coefficients for the non-Liouvillean process to be investigated.

In this article, we refer to the work of Jansen\cite{jansen},
who calculated these coefficients for the effect of
intra-beam scattering for charged particle beams
near thermodynamical equilibrium.
Including collision effects is particularly important
for highly charged heavy ions and high phase space densities.
We restrict ourselves to cases where the effect of collisions is small
compared with external forces or the smooth part of the self-fields.
Hence we do not treat the beam's turbulent heating phase, but the
long-term behavior of `relaxed' beams undergoing intra-beam scattering,
which typically occurs for beams circulating in storage rings.

The scalar collision model used below is a near-equilibrium closure,
not a complete theory of intra-beam scattering in an arbitrary lattice.
In particular, a dispersive storage ring requires additional
position--momentum correlations that are not retained in this closure.

Prior to mathematical details, we discuss in section~\ref{fopla} the
restrictions under which this Fokker-Planck
approach is applicable to charged particle beams.
Proceeding further, we will set up the extended
moment equations for the frame of reference moving with the beam
(section~\ref{moments}).
These moment equations are transformed to the
laboratory system in section~\ref{labor}, using the
usual `trace space' notation.
We thus obtain the extended envelope and emittance equations
(section~\ref{rms-eq}).
After discussing the concepts of `excess field energy'
and `temperature' of ion beams in sections~\ref{excess}
and~\ref{temperature}, respectively, we review in section~\ref{coeff}
the relationship between the scattering-related Fokker-Planck
coefficients and the beam parameters.
In the following, we derive differential
equations that describe the change of beam emittances
due to a temperature balancing process.
This is performed for unbunched beams in section~\ref{unbunched}
and for the more general cases of bunched beams and coasting beams
in storage rings in section~\ref{bunched} and~\ref{coasting}.
Finally, we will discuss some numerical examples
in section~\ref{numeric}.
The implications of our theory for beam transport devices
such as quadrupole channels and storage rings will be outlined.
In addition, a Fokker-Planck description is considered as a possible
model for numerical emittance growth in particle simulations.
Its coefficients must be determined from the numerical error process.
They need not satisfy the self-collision Einstein relation.
\section{THE FOKKER-PLANCK APPROACH FOR CHARGED PARTICLE BEAMS} \label{fopla}
The 6-dimensional phase space distribution function
$f = f(\vec{x},\vec{p};t)$ that represents a charged
particle beam satisfies Liouville's theorem
\begin{equation} \label{liouville}
\frac{d f}{d t} = 0 \; ,
\end{equation}
along the characteristics of a collisionless Hamiltonian mean-field model.
For charged particle beams whose electric self-fields
must be taken into account, Eq.~(\ref{liouville}) remains
true if the space charge potential $V^{\mathrm{sc}}$ is
sufficiently smooth that it can be treated analogous to
an external defocusing potential.

Writing (\ref{liouville}) explicitly and inserting the
equation of motion of a particle with the mass $m$ and charge $q$
under the influence of external focusing fields $\vec{F}^{\mathrm{ext}}$
and the smooth part of the self-fields $\vec{E}^{\mathrm{sc}}$
$$
\dot{\vec{p}} = \vec{F}^{\mathrm{ext}} + q\vec{E}^{\mathrm{sc}} \; ,
$$
we directly obtain the Vlasov equation.
The Poisson equation describes the
relationship between the distribution function $f$ and
the space-charge field $\vec{E}^{\mathrm{sc}}$:
$$
\div\vec{E}^{\mathrm{sc}} = 4\pi Nq \int f \; d^3 p \;.
$$
Gaussian electromagnetic units are used throughout.
The distribution is normalized by $\int f\,d^3x\,d^3p=1$ and $N$ is the
number of particles in the bunch or in the chosen periodic cell.
For an infinitely long coasting beam, a finite periodic cell or a
normalization per unit length must be specified instead.
The electric field is defined by the force $q\vec E$.

Collisional correlations, residual-gas scattering, radiation damping,
and cooling can require additional terms in the one-particle equation.
Their effect is not characterized solely by the geometrical support of $f$.
In particular, an increase of rms-emittance need not imply an increase
of fine-grained phase-space volume.
The term `non-Liouvillean' will refer below to the reduced one-particle
operator, not to the microscopic Hamiltonian dynamics of an isolated beam.

The action of non-Liouvillean effects can be described
mathematically replacing Eq.~(\ref{liouville})
by the Boltzmann equation:
$$
\frac{df}{dt} = \left[ \frac{\partial f}
{\partial t}\right]_{\mathrm{NL}} \;.
$$
Explicitly, this equation can be written in
its non-relativistic form as:
\begin{equation} \label{bz1}
\frac{\partial f}{\partial t} + \frac{\vec{p}}{m} \cdot
\vec{\nabla}_x f + \left( \vec{F}^{\mathrm{ext}} +
q\vec{E}^{\mathrm{sc}} \right) \cdot \vec{\nabla}_p f =
\left[ \frac{\partial f}{\partial t}\right]_{\mathrm{NL}} \;.
\end{equation}
The right-hand side describes the additional drift, diffusion, or collision
terms in the reduced one-particle distribution.
If this term can be neglected, Eq.~(\ref{bz1}) reduces to
the Vlasov equation.

Under certain conditions to be discussed below, the
non-Liouvillean effects can be modeled by the
Fokker-Planck equation (see References.~\cite{jansen,risken}, for example):
$$
\left[ \frac{\partial f}{\partial t} \right]_{\mathrm{NL}} =
- \sum_i \frac{\partial}{\partial p_i} \left\{ F_i(\vec{x},\vec{p},t) \, f \right\} +
m^2 \sum_{i,j} \frac{\partial^2}{\partial p_i \partial p_j}
\left\{ D_{ij}(\vec{x},\vec{p},t) \, f \right\} \;.
$$
Herein $D_{ij}$ stands for the `diffusion tensor' elements, whereas the
$F_i$ are referred to as the `drift vector' components.

The diffusion tensor is symmetric and positive semidefinite.
With the convention above, $2m^2D_{ij}$ is the conditional momentum
increment covariance per unit beam-frame time.
A diffusion approximation requires a separation of time scales
$$
\tau_{\mathrm{corr}}\ll\Delta t\ll\tau_{\mathrm{evol}} \;,
$$
where $\tau_{\mathrm{corr}}$ is the correlation time of the fluctuating force
and $\tau_{\mathrm{evol}}$ is the time on which the distribution changes appreciably.
Over such an interval, the accumulated momentum increments must be small
on the scale of the distribution, and higher jump moments must be negligible.
The Markov property alone does not imply a second-order Fokker-Planck equation.

The most difficult part in our task to include
non-Liouvillean effects in our analysis of beam optics
is to determine the Fokker-Planck coefficients
$D_{ij}$ and $F_i$.
Especially the effect of turbulent heating of the beam due
to mismatch -- where the beam is far from thermodynamical
equilibrium -- can make these coefficients
explicitly time dependent.
This makes it extremely complicated -- or even impossible --
to evaluate the Fokker-Planck coefficients on a rigorous basis,
hence forces to apply heuristic models\cite{bohn}.

Near an isotropic kinetic equilibrium, a scalar relaxation model can
simplify the moment equations.
Treating its coefficients as constant over a focusing period is a separate
approximation and is not guaranteed by proximity to equilibrium.
Density, temperature, and therefore the collision rate may vary along a ring.
When that variation matters, the rate must remain inside the period integrals.

Eq.~(\ref{bz1}) represents the self-consistent equation
of motion for a phase space distribution $f$.
It describes a charged particle beam under the
influence of external focusing forces,
the {\em smooth\/} part of the self-fields due to its
macroscopic charge density function, and the non-Liouvillean
effects of a Markov process within the beam, which
may change the distribution and its rms-emittances.
\section{THE EXTENDED MOMENT EQUATIONS} \label{moments}
In order to circumvent a lengthy direct integration
of Eq.~(\ref{bz1}), we switch from the total information
on the beam that is contained in the phase space distribution
function $f$ to physical quantities that contain
less information on the beam but still can serve as
measures for the beam quantities of interest.
A usual way to simplify our description of beam dynamics
is to derive equations of motion for the beam's second
central moments from the equation of motion for $f$.

Coordinates and momenta are measured relative to their centroids.
The displayed equations assume that these centroids remain zero under the
chosen symmetric forces.
Otherwise their equations must be subtracted when forming central moments.
With $f$ as the normalized particle phase space distribution function,
the beam moments can be defined by integrals
over all phase space, $\tau$:
\begin{eqnarray*}
\<x_i^2\>   & = & \int x_i^2   \, f \; d\tau \\
\<x_i p_i\> & = & \int x_i p_i \, f \; d\tau \\
\<p_i^2\>   & = & \int p_i^2   \, f \; d\tau \\
\<x_i E_i\> & = & \int x_i E_i \, f \; d\tau \\
\<p_i E_i\> & = & \int p_i E_i \, f \; d\tau \;.
\end{eqnarray*}
We obtain -- as an example -- the time derivative of $\< x_i^2 \>$ using
\begin{equation} \label{hderi}
\frac{d}{dt} \<x_i^2\> = \int x_i^2 \,
\frac{\partial   f}{\partial   t} d\tau
\end{equation}
and inserting $\partial   f / \partial   t$ from
Eq.~(\ref{bz1}) into (\ref{hderi}).
The following integrations by parts require the momentum fluxes and
weighted boundary terms to vanish at infinity, or appropriate no-flux
boundary conditions on a finite domain.
A singular idealization such as the K--V distribution is understood in a weak sense.
The non-vanishing moment terms assemble for each phase space plane
$i = 1,2,3$ to the following coupled set of equations:
\begin{eqnarray} \label{deri}
\frac{d}{dt} \<x_i^2\> - \frac{2}{m} \<x_i p_i\> & = & 0 \nonumber \\
\frac{d}{dt} \<x_i p_i\> - \frac{1}{m} \<p_i^2\> -
\<x_i F^{\mathrm{ext}}_i\> - q\, \<x_i E_i\> & = & \<x_i F_i \> \\
\frac{d}{dt} \<p_i^2\> - 2\, \<p_i F^{\mathrm{ext}}_i \> -
2q\, \<p_i E_i\> & = & 2\, \<p_i F_i \>+2m^2\< D_{ii} \> \nonumber \;.
\end{eqnarray}
The left-hand terms of Eqs.~(\ref{deri})
have been derived first by Sacherer\cite{sacherer},
assuming that Liouville's theorem applies.
The non-zero right-hand terms\cite{soerensen} originate in the
Fokker-Planck equation that describe the non-Liouvillean effects.
Only the diagonal diffusion coefficients enter the three selected
within-plane moment equations.
Off-diagonal coefficients enter cross-plane covariances and cannot be
discarded in a coupled six-dimensional description.
All terms of the right-hand side of (\ref{deri}) are
proportional to the Fokker-Planck coefficients.
As a consequence, they vanish
if the non-Liouvillean effects can be neglected.
The set of coupled differential equations
then reduces to Sacherer's equations, as expected.
\section{TRANSFORMATION TO THE LABORATORY SYSTEM} \label{labor}
For circular structures, the dependence of revolution time on momentum
is described by the phase-slip factor $\eta$.
An averaged longitudinal equation can be written using the effective coefficient
$$
m_{z;b}=\frac{m}{\eta\gamma^2} \;,
$$
with
\begin{equation} \label{etadef}
\eta = \cases{\gamma^{-2}-\gamma_t^{-2} &for a circular orbit
\cr \noalign{\vskip3mm}
\gamma^{-2} &for straight motion.}
\end{equation}
Here $\gamma_t$ denotes the transition $\gamma$.
This effective mass parametrizes phase-slip kinematics.
It is not the physical mass entering the local random kinetic energy or
the momentum diffusion coefficient.
The sign of $\eta$ must be retained in the longitudinal orbital dynamics.
We assume constant reference $\beta$, $\gamma$, and $\eta$.
If $\eta$ varies, differentiating $z'=\eta\delta$ generates additional
terms proportional to $\eta'/\eta$, and the displayed equations require modification.
A global inertial beam frame does not exist around a curved ring.
For a ring, the longitudinal formulas below are an averaged,
zero-dispersion reduction of local collision kicks, not a complete
Lorentz transformation of the ring dynamics.

Usually we may formulate the equation of motion
(\ref{bz1}) in the non-relativistic limit,
if we refer to the beam's center of momentum frame.
In this paper, we will restrict ourselves to these cases.
On the other hand, we are interested in the beam
properties as they occur in the laboratory system.
We therefore must transform all quantities from the beam frame
(index $b$) into the laboratory system (index $\ell$).
In the latter system, the time scale is `stretched' by
the Lorentz factor $\gamma$ compared to
the proper time of the reference particle.
Following the usual `trace space'\cite{lawson}
formulation of beam optics, we replace the proper time $t$
as the independent variable by the distance
$s$ along the focusing structure measured in the laboratory system:
$$
s=c\beta t_\ell=c\beta\gamma t_b \;, \qquad t\equiv t_b \;.
$$
For a straight reference orbit, the paraxial coordinate transformation is
$$
x_1 \equiv x_b = x_\ell \qquad , \qquad
x_2 \equiv y_b = y_\ell \qquad , \qquad
x_3 \equiv z_b = \gamma \, (\Delta z)_\ell \equiv \gamma z_\ell\; .
$$
For the momentum variables we use
$$
p_1 \equiv p_{x;b} = m c \beta \gamma \cdot x'_\ell \quad , \quad
p_2 \equiv p_{y;b} = m c \beta \gamma \cdot y'_\ell \quad , \quad
p_3 \equiv p_{z;b} = m c \beta \eta^{-1} z'_\ell \;.
$$
Here $\delta=\Delta p/p$ and
$$
p_{z;b}=mc\beta\delta,\qquad z'=\eta\delta \;.
$$
For a ring, $z$ denotes the longitudinal phase coordinate in the averaged
orbital equations, whereas $p_{z;b}$ is the local physical momentum deviation.
The relation $z_b=\gamma z$ is used as a spatial Lorentz relation only
for a straight beam, not for a dispersive closed orbit.
For quasistatic beam self-fields, the electric-field components satisfy
$$
E_{x;b} = \gamma^{-1} E_{x;\ell} \qquad , \qquad
E_{y;b} = \gamma^{-1} E_{y;\ell} \qquad , \qquad
E_{z;b} = E_{z;\ell} \;.
$$
First consider prescribed drag toward zero momentum in the reference frame.
This is an external-bath or test-particle model.
Its drift coefficients have the Stokes form
$$
F_i = -\beta_{f;i} \cdot p_i \; .
$$
The nonnegative $\beta_{f;i}(t_b)$ are beam-frame friction rates and
are taken to be spatially uniform in these moment equations.
External magnetic as well as electric focusing is included separately
in the prescribed orbital focusing functions.
Suppressing laboratory subscripts, we use the signed coefficients
$k_i^2(s)$ for the prescribed linear orbital focusing.
The superscript 2 is conventional notation and does not impose positivity
in a defocusing element.
For the prescribed-drag model, the transverse moment equations become
\begin{eqnarray} \label{deri2}
\frac{d}{ds} \<x^2\> - \, 2\<xx'\> & = & 0 \nonumber \\
\frac{d}{ds} \<xx'\> - \<x'^2\> + \, k_x^2(s) \<x^2\> -
\frac{q}{m c^2 \beta^2 \gamma^3} \<x E_x\> & = &
-\frac{\beta_{f;x}}{c \beta \gamma} \, \<xx'\> \\
\frac{d}{ds} \<x'^2\> + \, 2 k_x^2(s) \<xx'\> -
\frac{2q}{m c^2 \beta^2 \gamma^3} \<x' E_x\> & = &
-\frac{2\beta_{f;x}}{c \beta \gamma} \, \<x'^2\> +
\frac{2\< D_{xx}\>}{c^3 \beta^3 \gamma^3} \nonumber \;.
\end{eqnarray}
The corresponding set of equations holds for the $y$-direction.
For the $z$-direction, we obtain:
\begin{eqnarray} \label{deri3}
\frac{d}{ds} \<z^2\> - \, 2\<zz'\> & = & 0 \nonumber \\
\frac{d}{ds} \<zz'\> - \<z'^2\> + \, k_z^2(s) \<z^2\> -
\frac{q\cdot \eta}{m c^2 \beta^2 \gamma} \<z E_z\> & = &
-\frac{\beta_{f;z}}{c \beta \gamma} \, \<zz'\> \\
\frac{d}{ds} \<z'^2\> + \, 2 k_z^2(s) \<zz'\> -
\frac{2q\cdot \eta}{m c^2 \beta^2 \gamma} \<z' E_z\> & = &
-\frac{2\beta_{f;z}}{c \beta \gamma} \, \<z'^2\> +
\frac{2\eta^2\< D_{zz}\>}{c^3 \beta^3 \gamma} \nonumber \;.
\end{eqnarray}
The longitudinal diffusion term follows directly from
$z'=\eta p_{z;b}/(mc\beta)$ and $ds/dt_b=c\beta\gamma$.
It reduces to $2\<D_{zz}\>/(c^3\beta^3\gamma^5)$ only for straight motion.
The stochastic equations remain compatible with single-particle
Langevin equations, although not with deterministic trajectories alone.
For example, their beam-frame momentum increments can be represented by
$dp_i=\cdots+F_i\,dt_b+mB_{ij}\,dW_j$, with $BB^{\rm T}=2D$.

For elastic self-collisions, the collision operator must instead conserve
particle number, momentum, and kinetic energy at each spatial position.
A convenient reduced, conserving model is\cite{conservfp}
$$
C[f]=\beta_f\sum_i\frac{\partial}{\partial p_i}
\left[(p_i-\bar p_i)f+mkT_b(\vec x,t_b)
\frac{\partial f}{\partial p_i}\right] \;,
$$
where
$$
\bar p_i(\vec x,t_b)=\frac{\int p_i f\,d^3p}{\int f\,d^3p},\qquad
3mkT_b(\vec x,t_b)=
\frac{\int |\vec p-\bar{\vec p}|^2f\,d^3p}{\int f\,d^3p} \;.
$$
With a spatially uniform effective $\beta_f$, integration by parts gives
$$
\<x_iF_i\>=0,\qquad
\<p_iF_i\>=-\beta_f\<(p_i-\bar p_i)^2\> \;.
$$
The sum of the three collisional kinetic-energy changes vanishes because
$\<D_{ii}\>=\beta_f k\<T_b(\vec x,t_b)\>/m$.
This model represents isotropization with a scalar rate and does not
reproduce the full velocity dependence of Coulomb friction and diffusion.

In the subsequent second-moment closure, we approximate the full local
mean flow as affine and uncoupled,
$$
\bar p_i(\vec x,t_b)=
\frac{\<x_ip_i\>}{\<x_i^2\>}x_i \;.
$$
For this self-collision specialization, the right-hand side of the
second equation of (\ref{deri2}) is zero.
In its third equation, the drag contribution becomes
$$
-\frac{2\beta_f}{c\beta\gamma}
\left(\<x'^2\>-\frac{\<xx'\>^2}{\<x^2\>}\right) \;.
$$
The same replacements apply to (\ref{deri3}), using the corrected
longitudinal diffusion coefficient.
Thus coherent flow is not directly damped by the local self-collision operator.
The affine-flow approximation is a moment closure whose accuracy must
be checked for the distribution under consideration.

\section{THE EXTENDED ENVELOPE AND RMS-EMITTANCE EQUATIONS} \label{rms-eq}
Following further Sacherer's formalism, we define
the rms-emittance $\varepsilon_x(s)$ as
\begin{equation} \label{epsrms}
\varepsilon_x^2(s) \, = \,  \<x^2\> \<x'^2\> - \<x x'\>^2
\end{equation}
and first combine the two prescribed-drag equations at the beginning of
(\ref{deri2}) in order to set up a differential equation
for $\sqrt{\<x^2\>}$.
This quantity is of particular importance in the study of charged
particle beams, since it is proportional to the actual
beam width in the $x$-direction:
\begin{equation} \label{kv1}
\frac{d^2}{ds^2} \sqrt{\<x^2\>} + \frac{\beta_{f;x}}{c \beta \gamma}
\frac{d}{ds} \sqrt{\<x^2\>} + k_x^2(s) \sqrt{\<x^2\>} -
\frac{q}{m c^2 \beta^2 \gamma^3} \frac{\<xE_x\>}{\sqrt{\<x^2\>}} -
\frac{\varepsilon_x^2(s)}{\sqrt{\<x^2\>^3}} = 0 \;.
\end{equation}
Comparing Eq.~(\ref{kv1}) with Sacherer's
`rms envelope equation'\cite{sacherer}, we observe that
one additional term containing the first order
derivative of $\sqrt{\<x^2\>}$ appears in our theory.

For a prescribed bath, the term proportional to $\beta_{f;x}$
is a drag term.
It does not by itself prove stability or asymptotic matching in a
periodically driven, self-consistent envelope system.
For a stable, underdamped linear mode with a constant restoring
coefficient and negligible diffusion, the bath-drag model gives
$$
\tau_{f;b}=2/\beta_f,\qquad \tau_{f;\ell}=2\gamma/\beta_f \;.
$$
These times are not universal self-matching times for intra-beam scattering.

For the conserving self-collision specialization described in
section~\ref{labor}, the envelope equation instead reads
$$
\frac{d^2}{ds^2}\sqrt{\<x^2\>}+k_x^2(s)\sqrt{\<x^2\>}
-\frac{q}{mc^2\beta^2\gamma^3}
\frac{\<xE_x\>}{\sqrt{\<x^2\>}}
-\frac{\varepsilon_x^2}{\sqrt{\<x^2\>^3}}=0 \;.
$$
Collisions influence the envelope through the evolving pressure and
emittance, not through a direct force on the local mean flow.
Collective damping, when present, requires a consistent dynamical calculation.

In general, $\<xE_x\>$ and $\varepsilon_x^2$ are not
constants of motion but unknown functions of $s$.
This accounts for the fact that Eq.~(\ref{kv1})
is not closed, hence cannot be used directly
to evaluate the envelope function.
In the following, we will discuss the conditions
under which Eq.~(\ref{kv1}) can be combined
with additional differential equations in order to
establish a closed set of equations that can be integrated.

On the basis of Eqs.~(\ref{deri2}), the derivative of the
rms-emittance (\ref{epsrms}) is readily calculated to give
\begin{equation} \label{epsderi1}
\frac{1}{\<x^2\>} \frac{d}{ds} \varepsilon_x^2(s) =
\frac{2q}{m c^2 \beta^2 \gamma^3}
\left( \<x'E_x\> - \frac{\<xx'\>}{\<x^2\>} \<xE_x\>\right) -
2\left( \frac{\beta_{f;x}}{c \beta \gamma} \frac{\varepsilon_x^2(s)}{\<x^2\>} -
\frac{\< D_{xx}\>}{c^3 \beta^3 \gamma^3} \right) \;.
\end{equation}
For affine local mean flow, the conserving self-collision model gives
the same emittance equation, although its envelope equation has no bath-drag term.
The longitudinal emittance equation is
$$
\frac{1}{\<z^2\>} \frac{d}{ds} \varepsilon_z^2(s) =
\frac{2q\eta}{m c^2 \beta^2 \gamma}
\left( \<z'E_z\> - \frac{\<zz'\>}{\<z^2\>} \<zE_z\>\right) -
2\left( \frac{\beta_{f;z}}{c \beta \gamma} \frac{\varepsilon_z^2(s)}{\<z^2\>} -
\frac{\eta^2\< D_{zz}\>}{c^3 \beta^3 \gamma} \right) \;.
$$
In the following section, we analyze the physical meaning
of those moments that contain electric field components $E_i$.
Thereafter, the terms related to the Fokker-Planck
coefficients will be investigated.
We will show that they describe the relaxation
of temperature differences within the beam.
\section{THE `EXCESS FIELD ENERGY' OF A CHARGED PARTICLE BEAM} \label{excess}
As has been discussed previously\cite{wang,host,stho}, field-energy
changes can be related to moments of the self-force.
All quantities in the following energy identities are evaluated in one
inertial beam frame, within the electrostatic approximation.
For a finite isolated bunch, and with no boundary energy flux,
\begin{equation} \label{dw}
\frac{dW_b}{dt_b}=-\frac{Nq}{m}\sum_i\<p_{i;b}E_{i;b}\> \;.
\end{equation}
For a longitudinally uniform beam, the corresponding energy per unit
length and a common transverse boundary convention must be used.
These identities are not global inertial-frame energy identities for a ring.

Choose a uniform ellipsoidal reference distribution with the same charge
and instantaneous rms-dimensions as the actual beam.
Its interior electric field is linear in the coordinates.
Its field energy satisfies
$$
\frac{dW_{u;b}}{dt_b}=-Nq\sum_i
\frac{\<x_{i;b}p_{i;b}\>}{m\<x_{i;b}^2\>}
\<x_{i;b}E_{i;b}\>_u \;.
$$
The subscript $u$ refers to the reference distribution.
Its dimensions and their derivatives agree with those of the actual beam,
but its force moments need not agree.
Consequently,
\begin{equation} \label{dwdwu}
\frac{d(W_b-W_{u;b})}{dt_b}=-Nq\sum_i\left[
\frac{\<p_{i;b}E_{i;b}\>}{m}
-\frac{\<x_{i;b}p_{i;b}\>}{m\<x_{i;b}^2\>}
\<x_{i;b}E_{i;b}\>_u\right] \;.
\end{equation}
In particular, the reference force moment must not be replaced by the
actual force moment merely because the two distributions have equal rms-sizes.
For some transversely elliptic, unbunched comparison families these
moments agree, but that is not a general three-dimensional property.

The directional moments appearing in the emittance equations are, in laboratory notation,
$$
M_i=\<x_i'E_i\>
-\frac{\<x_ix_i'\>}{\<x_i^2\>}\<x_iE_i\> \;.
$$
In the beam frame, define the analogous velocity--field residuals
$$
M_{i;b}=\frac{\<p_{i;b}E_{i;b}\>}{m}
-\frac{\<x_{i;b}p_{i;b}\>}{m\<x_{i;b}^2\>}
\<x_{i;b}E_{i;b}\> \;.
$$
Equation~(\ref{dwdwu}) then implies
$$
\sum_iM_{i;b}=-\frac{1}{Nq}\frac{d(W_b-W_{u;b})}{dt_b}
-\sum_i\frac{\<x_{i;b}p_{i;b}\>}{m\<x_{i;b}^2\>}
\left(\<x_{i;b}E_{i;b}\>-\<x_{i;b}E_{i;b}\>_u\right) \;.
$$
This additional reference-moment term is required for a general bunch.
For example, under a nontrivial homothetic expansion of a nonuniform
three-dimensional bunch, every $M_{i;b}$ can vanish while both Coulomb
energies scale inversely with the size and $W_b-W_{u;b}$ changes.

Within comparison families for which the uniform reference has minimum
field energy at fixed rms-dimensions\cite{host}, $W_b-W_{u;b}$ is a
nonnegative `excess' energy.
Its conversion into random kinetic energy depends on the accompanying
size evolution and external work.
It is not by itself a general bound on emittance growth in a driven lattice.
Relaxation of a density profile can occur within a few plasma
periods in the examples discussed previously\cite{strekla,wang,anderson}.
Nevertheless, neglecting each $M_i$ in a long-term emittance equation is
an additional directional closure assumption.
A constant excess energy, or a vanishing sum of residuals, does not imply
that every directional residual vanishes.

\section{THE TEMPERATURE OF A CHARGED PARTICLE BEAM} \label{temperature}
As usual in statistical physics, we relate the temperature
of an ensemble of particles to their `incoherent' motion.
Since charged particle beams change their size
while passing through a focusing device,
we must subtract the `coherent' part of the particle
velocities from the total velocities in order to obtain
physically reasonable expressions for the beam temperatures.

The temperatures in all spatial directions are not necessarily the same.
We thus have to define them separately for each direction.
In order to determine the temperature $T_x$,
we must first calculate $g(x,x';s)$ denoting
the projection of the phase space distribution
function $f$ onto the $x,x'$-subspace at $s$:
$$
g(x,x';s) = \int f \, dy \, dy' \, dz \, dz' \;.
$$
Here $f$ denotes the normalized trace-space density, including the
Jacobian of the momentum-to-slope transformation.
The scaled `coherent' velocity $\xi'(x;s)$ in the $x$-direction
as a function of $x$ is given by:
\begin{equation} \label{xiprime}
\xi'(x;s) = \frac{\int x' g(x,x';s) \; dx'}{\int g(x,x';s) \; dx'} \;.
\end{equation}
Now the laboratory $x$-temperature $kT_x$ in energy units
can be defined as\cite{lawson}:
$$
k T_x = m c^2 \beta^2 \gamma \int (x' - \xi')^2 \, f \; d\tau \; ,
$$
which is easily simplified to
\begin{equation} \label{temp-x}
k T_x = m c^2 \beta^2 \gamma \Bigl( \< x'^2 \> - \< \xi'^2 \> \Bigr) \;.
\end{equation}
In general, it is not possible to calculate
$\xi'(x;s)$ for a given distribution function $f$
in a closed analytical form.
As an example that can be treated analytically,
we quote the `K-V' model\cite{kapvla}
for the distribution function $f$ of an unbunched beam.
Its transverse four-dimensional distribution has uniformly populated
nondegenerate two-dimensional projections.
For this case, the integrals contained in (\ref{xiprime})
are easily solved to give:
\begin{equation} \label{xiprimekv}
\xi'(x;s) = \frac{\< x x' \>}{\< x^2 \>} \, x \qquad \leadsto \qquad
\< \xi'^2 \> = \frac{\< x x' \>^2}{\< x^2 \>} \;.
\end{equation}
Equation~(\ref{xiprimekv}) also applies to other distributions whose
conditional mean velocity is affine in $x$, including an uncoupled Gaussian.
For a general distribution, define
$$
a_x=\frac{\<xx'\>}{\<x^2\>},\qquad
kT_x^{\mathrm{rms}}=mc^2\beta^2\gamma
\frac{\varepsilon_x^2}{\<x^2\>} \;.
$$
The exact error identity is
$$
k(T_x^{\mathrm{rms}}-T_x)
=mc^2\beta^2\gamma\<(\xi'(x;s)-a_xx)^2\>\geq0 \;.
$$
The rms formula is therefore an upper bound on the conditional-variance
temperature and becomes exact for affine conditional mean flow.
An extremum of the beam width gives $\<xx'\>=0$ but does not imply
$\xi'(x;s)=0$.
There is no general cancellation of the temperature error at such an extremum.
For the self-collision closure in section~\ref{labor}, the stronger
condition on the mean momentum at fixed full position $\vec x$ is required.
In particular, coupled spatial flows must not be hidden by taking a
one-coordinate marginal.

Under the affine, uncoupled mean-flow approximation, we use the
rms-based temperatures in the remaining moment equations.
The transverse laboratory-frame kinetic-energy convention is
$$
kT_x=\frac{\<(p_{x;\ell}-\bar p_{x;\ell})^2\>}{m\gamma}
=mc^2\beta^2\gamma\frac{\varepsilon_x^2}{\<x^2\>} \;.
$$
This defines a kinetic temperature proxy with $T_{i;b}=\gamma T_{i;\ell}$.
It is not a general relativistic transformation law for thermodynamic temperature.
The physical longitudinal random momentum is
$p_{z;b}=mc\beta\delta=mc\beta z'/\eta$.
Thus, for constant nonzero $\eta$,
$$
kT_z=\frac{mc^2\beta^2}{\gamma\eta^2}
\frac{\varepsilon_z^2}{\<z^2\>}
=mc^2\beta^2\gamma\frac{\bar\varepsilon_z^2}{\<z^2\>},
\qquad
\bar\varepsilon_z=\frac{\varepsilon_z}{\gamma|\eta|} \;.
$$
The bar now denotes a kinetic rescaling of longitudinal emittance.
For straight motion this gives $\bar\varepsilon_z=\gamma\varepsilon_z$.
For a ring the physical collision temperature must not be calculated
from the signed effective longitudinal mass.
An equivalent, nonsingular definition in momentum coordinates is
$$
\bar\varepsilon_z^2=\gamma^{-2}
\left[\<z^2\>\<\delta^2\>-\<z\delta\>^2\right] \;.
$$
This is the definition to use at transition, where the map from $\delta$
to $z'$ degenerates.
The square in the random kinetic energy makes the temperature positive.
It does not imply a symmetry of the complete ring dynamics above and below transition.

The equilibrium temperature $T$ in the scalar relaxation model is the
average of the three kinetic temperatures.
When they are spatially nonuniform, this is their particle-weighted mean,
\begin{equation} \label{temp3}
\frac{k T}{m c^2 \beta^2 \gamma} = \frac{1}{3} \left(
\frac{\varepsilon_x^2(s)}{\<x^2\>} +
\frac{\varepsilon_y^2(s)}{\<y^2\>} +
\frac{\bar{\varepsilon}_z^2(s)}{\<z^2\>} \right) \;.
\end{equation}
If the longitudinal temperature need not to be taken into account, we
can restrict ourselves to the average of the transverse temperatures.
The transverse equilibrium temperature $T_\perp$ can then be defined as:
\begin{equation} \label{temp2}
\frac{k T_\perp}{m c^2 \beta^2 \gamma} = \frac{1}{2} \left(
\frac{\varepsilon_x^2(s)}{\<x^2\>} +
\frac{\varepsilon_y^2(s)}{\<y^2\>} \right) \;.
\end{equation}
Equation~(\ref{temp2}) defines a deliberately restricted transverse
relaxation model.
Its use requires negligible collisional energy exchange with the
longitudinal degree of freedom on the time scale of interest.
An unbunched beam in a non-dispersive channel does not satisfy that
requirement automatically.
The general local three-dimensional relaxation model uses (\ref{temp3}).
Dispersive ring dynamics additionally requires transverse--longitudinal
correlations beyond the uncoupled model considered here.

\section{THE FOKKER-PLANCK COEFFICIENTS FOR INTRA-BEAM SCATTERING}\label{coeff}
So far, the procedure of deriving extended moment equations
has been performed formally by speaking in general of
`non-Liouvillean' effects to be described by the Fokker-Planck equation.
In other words, up to now nothing has been said about
the Fokker-Planck coefficients $\beta_{f;i}$ and $D_{ii}$.
As an example requiring correlations beyond the collisionless mean-field equation,
we will treat the process of intra-beam scattering.
In order to learn how the Fokker-Planck coefficients for the effect
of intra-beam scattering are correlated to the physical beam properties,
it is necessary to study the process of Coulomb scattering of an
ensemble of particles in detail.
This has been done extensively by Jansen\cite{jansen},
treating the more general `many body problem of particles
interacting through an inverse square force law'.
We therefore can restrict ourselves to
results that apply in this context.

The method of evaluating the Fokker-Planck coefficients
for intra-beam scattering in a charged particle
beam can be summarized as follows:
\begin{itemize}
\item in the first step, the velocity changes of a test particle
due to scattering from a single beam particle as a function of the test
particle's initial velocity and impact parameter are calculated,
\item second, the expression obtained in the first step is
averaged over all possible impact parameters,
\item finally, averaging over all particle velocities
of the beam is performed.
This means that the velocity distribution of the beam
must be taken into account.
\end{itemize}
For simplicity, we assume that the {\em equilibrium\/} particle
beam has an {\em isotropic\/} Maxwellian velocity distribution.
For a distribution not too far from this state, we adopt the scalar
conserving relaxation closure of section~\ref{labor}.
This remains an approximation to Coulomb transport, whose exact
coefficients can depend on velocity and position even for an isotropic background.
Then only one diffusion coefficient $D$ in conjunction with a single
friction coefficient $\beta_f$ appears in our equations:
$$
D \equiv \<D_{xx}\> = \<D_{yy}\> = \<D_{zz}\> \qquad , \qquad
\beta_f \equiv \beta_{f;x} = \beta_{f;y} = \beta_{f;z} \;.
$$
For this closure, $D$ is the particle average of the local diffusion
coefficient and $T_b$ is the corresponding particle-averaged kinetic temperature.
With spatially uniform effective $\beta_f$, energy conservation gives
the Einstein\cite{einstein} relation
\begin{equation} \label{diffu}
D = \beta_f \cdot \frac{k T_b}{m} = \beta_f \cdot \gamma\frac{k T_\ell}{m}
\end{equation}
As a phenomenological near-equilibrium choice for the scalar rate,
we retain the nonrelativistic estimate associated with the original
Coulomb-scattering approximation\cite{jansen,reiser}, now written entirely
in beam-frame quantities,
\begin{equation} \label{friction}
\beta_f \simeq \frac{16\sqrt{\pi}}{3}\,n_b c
\left(\frac{q^2}{mc^2}\right)^2
\left(\frac{mc^2}{2kT_b}\right)^{3/2}\ln\Lambda \;.
\end{equation}
Here $n_b$ and $T_b$ are the beam-frame number density and kinetic
temperature, $m$ is the rest mass, and $q$ is the particle charge.
The numerical prefactor is a model choice, not a universal relaxation
rate for an anisotropic beam.
A quantitative application should determine the relevant energy-exchange
rate from the chosen Coulomb collision operator or from a calibrated model.
The laboratory rate is $\beta_{f;\ell}=\beta_f/\gamma$.
For a straight, uniformly moving beam, $n_\ell=\gamma n_b$ and the
kinetic-temperature convention of section~\ref{temperature} gives
$T_b=\gamma T_\ell$.
These conversions must be made before using laboratory inputs.

As usual for phenomena involving Coulomb scattering, we must establish
a reasonable upper limit for the maximum
impact parameter in order to keep $\Lambda$ finite.
For non-neutralized systems, it is suggested by Jansen\cite{jansen}
to identify the maximum impact parameter with the average distance
between the particles rather than with the Debye length.
This leads to the following formula for $\Lambda$:
$$
\Lambda\simeq\frac{3}{2}\frac{kT_b}{q^2 n_b^{1/3}} \;.
$$
This is a cutoff prescription for the stated weak-scattering model.
It requires $\Lambda\gg1$ and must be reconsidered when the beam size,
shielding, or correlation length imposes a different large-impact cutoff.
Logarithmic dependence reduces cutoff sensitivity only while the
weak-coupling approximation remains valid.
Using the conserving affine-flow specialization of
Eq.~(\ref{epsderi1}) and inserting (\ref{diffu}) yields
\begin{equation} \label{epsderi1a}
\frac{1}{\<x^2\>} \frac{d}{ds} \varepsilon_x^2(s) =
\frac{2q}{m c^2 \beta^2 \gamma^3}
\left( \<x'E_x\> - \frac{\<xx'\>}{\<x^2\>} \<xE_x\>\right) -
2 k_f \left( \frac{\varepsilon_x^2(s)}{\<x^2\>} -
\frac{kT}{m c^2 \beta^2 \gamma} \right) \; ,
\end{equation}
Here $k_f(s)=\beta_f(s)/(c\beta\gamma)$.
The temperature $T$ in the laboratory moment equations means $T_\ell$.
We observe that the term proportional to $k_f$
agrees with the temperature relaxation ansatz\cite{erlangen}
commonly used to describe temperature relaxation phenomena.

In the following sections, we insert the expressions (\ref{temp3}) and
(\ref{temp2}) for the equilibrium temperatures into Eq.~(\ref{epsderi1a}).
This will lead to fairly simple growth rate formulae suitable to
estimate the effect of intra-beam scattering on the beam emittance.
\section{THE EMITTANCE EQUATION FOR UNBUNCHED BEAMS} \label{unbunched}
Provided that collisional energy exchange between the longitudinal
and transverse degrees of freedom can be neglected, it is convenient to
define a `transverse equilibrium temperature' $T_\perp$
according to (\ref{temp2}).
For this reduced operator, the collision sum runs over $x$ and $y$,
and $2mkT_{\perp;b}$ replaces $3mkT_b$ in its local temperature definition.
We insert this expression into Eq.~(\ref{epsderi1a}):
\begin{equation} \label{epsderiunbu}
\frac{1}{\<x^2\>} \frac{d}{ds} \varepsilon_x^2(s) +
k_f \left(
\frac{\varepsilon_x^2(s)}{\<x^2\>} -
\frac{\varepsilon_y^2(s)}{\<y^2\>} \right) =
\frac{2q}{m c^2 \beta^2 \gamma^3}
\left( \<x'E_x\> - \frac{\<xx'\>}{\<x^2\>} \<xE_x\> \right) \;.
\end{equation}
The right-hand side contains the directional field residual $M_x$
defined in section~\ref{excess}.
For the closed relaxation model below, we neglect $M_x$ and $M_y$
separately, either identically within an affine-flow model or as an
explicit approximation to their secular contributions.
This is not a consequence of constant total excess field energy alone.

As has been shown by Sacherer\cite{sacherer},
the moment $\< x E_x \>$ contained in Eq.~(\ref{kv1})
can be expressed in terms of second order moments for an unbounded,
longitudinally uniform beam with elliptically stratified transverse density
and with no image fields,
\begin{equation} \label{ellisymm}
\< x E_x \> = \frac{I}{c\beta} \, \frac{\sqrt{\< x^2 \>}}
{\sqrt{\< x^2 \>} + \sqrt{\< y^2 \>}} \;.
\end{equation}
The elliptic-symmetry assumption can be useful for unbunched beams
in linear focusing channels, including profiles with nonlinear self-fields.
Its validity must be checked when the distribution changes shape.

With the help of (\ref{ellisymm}), we can
combine the self-collision envelope equation of section~\ref{rms-eq}
with Eq.~(\ref{epsderiunbu}) to
establish a closed set of coupled differential equations
for the `long-term' values of envelopes and emittances
containing solely second order beam moments:
\begin{eqnarray} \label{kv2}
\frac{1}{\< x^2 \>} \frac{d}{d s} \varepsilon_x^2(s) + k_f \left(
\frac{\varepsilon_x^2(s)}{\< x^2 \>} - \frac{\varepsilon_y^2(s)}{\< y^2 \>}
\right) & = & 0 \nonumber \\
\frac{1}{\< y^2 \>} \frac{d}{d s} \varepsilon_y^2(s) + k_f \left(
\frac{\varepsilon_y^2(s)}{\< y^2 \>} - \frac{\varepsilon_x^2(s)}{\< x^2 \>}
\right) & = & 0 \nonumber \\
\frac{d^2}{d s^2} \sqrt{\<x^2\>} +
k_x^2(s) \sqrt{\<x^2\>} -
\frac{K/2}{\sqrt{\<x^2\>} + \sqrt{\<y^2\>}} -
\frac{\varepsilon_x^2(s)}{\sqrt{\<x^2\>^3}} & = & 0 \nonumber \\
\frac{d^2}{d s^2} \sqrt{\<y^2\>} +
k_y^2(s) \sqrt{\<y^2\>} -
\frac{K/2}{\sqrt{\<x^2\>} + \sqrt{\<y^2\>}} -
\frac{\varepsilon_y^2(s)}{\sqrt{\<y^2\>^3}} & = & 0 \nonumber \\
& &
\end{eqnarray}
with $K$ denoting the dimensionless `generalized perveance'
$$
K = \frac{2qI}{m c^3 \beta^3 \gamma^3} \;.
$$
In order to gain a better insight into the mathematical
implications of these equations, we introduce $r_t$
as the ratio of the transverse beam temperatures
\begin{equation} \label{temprat}
r_t(s) = \frac{T_x}{T_y} = \frac{\varepsilon_x^2(s)}{\< x^2 \>}
\cdot  \frac{\< y^2 \>}{\varepsilon_y^2(s)} \; ,
\end{equation}
and rewrite the first two equations of (\ref{kv2})
in an equivalent form
\begin{eqnarray} \label{epsderi4}
\frac{d}{d s} \ln \varepsilon_x^2(s) & = & k_f \,
\bigl( r_t^{-1}(s) - 1 \bigr) \nonumber \\
\frac{d}{d s} \ln \varepsilon_y^2(s) & = & k_f \,
\bigl( r_t(s) - 1 \bigr) \;.
\end{eqnarray}
We observe that the rms-emittance in the $x$-direction
increases as long as the temperature $T_x$
is smaller than $T_y$.
At the same time, $\varepsilon_y$ decreases until a temperature
balance in both transverse directions is achieved.
These changes take place in the opposite
direction as long as $T_x > T_y$.
If we add Eqs.~(\ref{epsderi4}), we find a quantity
that increases in any case:
\begin{equation} \label{entro}
\frac{d}{d s} \ln\bigl[\varepsilon_x^2(s)\varepsilon_y^2(s)\bigr] =
k_f \cdot \frac{\bigl[ 1 - r_t(s) \bigr]^2}{r_t(s)} \geq 0 \;.
\end{equation}
For uncoupled transverse covariance blocks, the product
$\varepsilon_x^2\varepsilon_y^2$ is the determinant of the four-dimensional
transverse covariance matrix.
We define $\varepsilon_\perp=(\varepsilon_x\varepsilon_y)^{1/2}$.
Its increase in (\ref{entro}) is a result for the stated relaxation closure.
The logarithm of the emittance product, relative to a fixed reference,
is an rms entropy measure\cite{lalaglu}.
It equals the corresponding Gaussian entropy change, but is not the
fine-grained entropy of an arbitrary non-Gaussian distribution.
Cross-plane coupling requires the full covariance determinant instead
of a product of projected emittances.

We now formally integrate Eq.~(\ref{entro}):
\begin{equation} \label{epsint}
\frac{\varepsilon_x^2(s)\, \varepsilon_y^2(s)}
     {\varepsilon_x^2(0)\, \varepsilon_y^2(0)}
= \exp \left( \int\limits_0^s k_f(u)
\frac{[1-r_t(u)]^2}{r_t(u)}\,du \right)\geq1 \;.
\end{equation}
The inequality holds since $k_f$ as well as the
integrand cannot take on negative values.
It is not possible to solve Eq.~(\ref{epsint}) for
$\varepsilon_x\varepsilon_y$ since the integrand
itself is a function of the transverse emittances.
The explicit emittance ratio can be approximated by unity when
$$
\varepsilon_x(s) \approx \varepsilon_y(s) \;.
$$
This approximation requires an approximately equal-emittance solution,
for example in an appropriately symmetric lattice over the averaging interval.
It does not follow from temperature relaxation in an arbitrary lattice.
The envelopes can still depend implicitly on the changing emittances.
With this restriction in mind, the temperature ratio
(\ref{temprat}) can be replaced by the `envelope ratio':
$$
r_t(s) \approx r_e(s) = \frac{\< y^2 \>}{\< x^2 \>} \; ,
$$
yielding
$$
\varepsilon_\perp^4(s) = \varepsilon_\perp^4(0) \cdot \exp \left(
\int\limits_0^s k_f(u)
\frac{[1-r_e(u)]^2}{r_e(u)}\,du \right) \;.
$$
For periodic beam focusing systems, it is helpful to
introduce the following dimensionless quantity $I_e$:
\begin{equation} \label{envint}
I_e = \frac{1}{S} \int\limits_0^S
\frac{\bigl[ 1 - r_e(z) \bigr] ^2}{r_e(z)} \, dz \; ,
\end{equation}
with $S$ as the period length.
In the following sections, we will refer to $I_e$
as the `ellipticity' of the focusing system.
Since the integrand cannot take on negative
values, we find that $I_e \geq 0$.
The integral depends solely on the course of
the envelope ratio $r_e(s)$ which a matched beam
defines while passing through one focusing period.
In general it depends on both the lattice and the matched beam parameters,
including perveance and emittances.
Only under additional restrictions can it be regarded as a lattice-only quantity.

For a frozen matched solution and constant $\beta_f$, the corresponding
laboratory-time growth law is
$$
\frac{\varepsilon_\perp(t_\ell)}{\varepsilon_\perp(0)}
=\exp\left(\frac{\beta_f I_e}{4\gamma}t_\ell\right) \;.
$$
More generally, averaging the instantaneous rate over one period gives
\begin{equation} \label{e-fold}
\tau_{\mathrm{ef},\ell}^{-1}
=\frac{1}{4\gamma S}\int_0^S\beta_f(s)
\frac{[1-r_e(s)]^2}{r_e(s)}\,ds
\simeq\frac{\beta_f I_e}{4\gamma} \;.
\end{equation}
The final expression requires an effectively constant rate over that period.
For unequal emittances, $r_t$ replaces $r_e$ in the integral.
The beam-frame e-folding time is $\tau_{\mathrm{ef},\mathrm{b}}=\tau_{\mathrm{ef},\ell}/\gamma$.

We observe that the rate of the emittance growth due to
intra-beam scattering depends, besides $\beta_f$, on the
`ellipticity' $I_e$ of the beam envelopes along the focusing structure.
If the temperature imbalance persists on a set of nonzero length where
$\beta_f>0$, the reduced model predicts a positive instantaneous growth
rate of the emittance product.
Ordinary alternating-gradient quadrupole channels generally produce such
an imbalance, but this is not a theorem for every focusing device.

Note that (\ref{e-fold}) is valid only as long as the density, temperature, and matched-envelope ratios
are not appreciably changed on the growth interval.
This means that Eq.~(\ref{e-fold}) provides us with the
{\em instantaneous\/} emittance growth rate for given beam parameters.

In section~\ref{numeric}, we will present
the results of numerical calculations.
Of course, the effect of emittance growth due to
weak particle-particle collisions can be neglected in a particular
straight transport structure when its traversal time is small compared
with the laboratory growth time.
This is a time-scale condition rather than a general property of straight structures.
Nevertheless, the relation (\ref{e-fold}) gives us an insight
into the physical consequences of processes that involve statistically
fluctuating forces -- as they also occur due to simplifications
in computer simulations of charged particle beams.
We will shortly touch this subject in the last section.
\section{THE EMITTANCE EQUATION FOR BUNCHED BEAMS} \label{bunched}
For the uncoupled three-dimensional relaxation closure, we insert
Eq.~(\ref{temp3}) into Eq.~(\ref{epsderi1a}).
The definitions of $\bar\varepsilon_z$ and kinetic temperature are those
of section~\ref{temperature}.
We obtain
\begin{equation} \label{epsderi2}
\frac{1}{\<x^2\>} \frac{d}{ds} \varepsilon_x^2(s) =
\frac{2q}{m c^2 \beta^2 \gamma^3} \!
\left( \! \<x'E_x\> - \frac{\<xx'\>}{\<x^2\>} \<xE_x\> \! \right) -
\frac{2k_f}{3} \! \left( \!
\frac{2\varepsilon_x^2(s)}{\<x^2\>} -
\frac{\varepsilon_y^2(s)}{\<y^2\>} -
\frac{\bar{\varepsilon}_z^2(s)}{\<z^2\>} \! \right)
\end{equation}
The equation for $\varepsilon_y^2$ follows by interchanging $x$ and $y$.
For the rescaled longitudinal emittance the explicit equation is
$$
\begin{array}{rcl}
\displaystyle\frac{1}{\<z^2\>}\frac{d\bar\varepsilon_z^2}{ds}
&=&\displaystyle\frac{2q}{mc^2\beta^2\gamma^3\eta}M_z\\[2mm]
&&\displaystyle-\frac{2k_f}{3}\left(
\frac{2\bar\varepsilon_z^2}{\<z^2\>}
-\frac{\varepsilon_x^2}{\<x^2\>}
-\frac{\varepsilon_y^2}{\<y^2\>}\right) \;.
\end{array}
$$
This follows from the $\eta^2$ diffusion term in (\ref{deri3}) and
$\bar\varepsilon_z^2=\varepsilon_z^2/(\gamma^2\eta^2)$.
It applies for nonzero constant $\eta$ within the stated orbital approximation.
At transition the momentum-coordinate equations must be used instead.
Adding the three equations gives the moment balance
\begin{equation} \label{sr2}
\frac{1}{\<x^2\>}\frac{d\varepsilon_x^2}{ds}
+\frac{1}{\<y^2\>}\frac{d\varepsilon_y^2}{ds}
+\frac{1}{\<z^2\>}\frac{d\bar\varepsilon_z^2}{ds}
=\frac{2q}{mc^2\beta^2\gamma^3}
\left(M_x+M_y+\frac{M_z}{\eta}\right) \;.
\end{equation}
The collision terms cancel by kinetic-energy conservation.
The field residuals have their actual transformation weights.
They cannot in general be replaced by an unweighted derivative of
$W-W_u$.
For a straight beam, their relation to the electrostatic beam-frame
energy must also include the reference-moment term in section~\ref{excess}.

If each directional field residual is negligible in the intended
long-term approximation, Eq.~(\ref{epsderi2}) reduces to
\begin{equation} \label{epsderi3}
\frac{1}{\<x^2\>} \frac{d}{ds} \varepsilon_x^2(s) =
-\frac{2k_f}{3} \left(
\frac{2\varepsilon_x^2(s)}{\<x^2\>} -
\frac{\varepsilon_y^2(s)}{\<y^2\>} -
\frac{\bar{\varepsilon}_z^2(s)}{\<z^2\>} \right) \; ,
\end{equation}
again likewise for the $y$- and $z$-directions.

Together with the self-collision envelope equations, without the
bath-drag term of (\ref{kv1}), these equations form a closed set only
after specifying the field moments for a chosen charge-density family.
For example, a uniform ellipsoidal model supplies such moments\cite{sacherer}.
The diagonal system omits transverse--longitudinal coupling and is not
a complete bunched-beam model for a dispersive ring.

Eq.~(\ref{epsderi3}) can be written in an equivalent form
in terms of temperature ratios, as defined in Eq.~(\ref{temprat})
\begin{equation} \label{lnepsderi}
\frac{d}{d s} \ln \varepsilon_x^2(s) = \frac{2k_f}{3} \, \left(
\frac{T_y(s)}{T_x(s)} + \frac{T_z(s)}{T_x(s)} - 2 \right) \; ,
\end{equation}
and similarly for $\varepsilon_y^2$ and $\bar{\varepsilon}_z^2$.
Depending on the actual temperature ratios, the expression
in parentheses can be positive or negative.
This means that according to the direction of the temperature
flow within the beam, the rms-emittance $\varepsilon_x^2(s)$
can increase or decrease.
Within this closure, the product
$\varepsilon_x^2\varepsilon_y^2\bar{\varepsilon}_z^2$ cannot decrease
during the balancing process.
This is verified, if we add (\ref{lnepsderi}) to the
similar equations for $\varepsilon_y^2$ and $\bar{\varepsilon}_z^2$:
\begin{equation} \label{entro3}
\frac{d}{d s} \ln\bigl[\varepsilon_x^2(s)
\varepsilon_y^2(s)\bar{\varepsilon}_z^2(s)\bigr] =
\frac{2k_f}{3} \, \left(
\frac{(1-r_{xy})^2}{r_{xy}} +
\frac{(1-r_{xz})^2}{r_{xz}} +
\frac{(1-r_{yz})^2}{r_{yz}} \right) \geq 0 \;.
\end{equation}
For the sake of clarity,
we used abbreviations for the temperature ratios, namely
$$
r_{xy}(s) = \frac{T_y(s)}{T_x(s)} \qquad , \qquad
r_{xz}(s) = \frac{T_z(s)}{T_x(s)} \qquad , \qquad
r_{yz}(s) = \frac{T_z(s)}{T_y(s)} \;.
$$
As already demonstrated in the previous section,
Eq.~(\ref{entro3}) can be formally integrated:
\begin{equation} \label{t-growth}
\frac{\varepsilon_{\mathrm{tot}}^3(S)}{\varepsilon_{\mathrm{tot}}^3(0)} \equiv
\frac{\varepsilon_x(S) \, \varepsilon_y(S) \, \bar{\varepsilon}_z(S)}
{\varepsilon_x(0) \, \varepsilon_y(0) \, \bar{\varepsilon}_z(0)}
=\exp\left\{\frac{1}{3}\int_0^S k_f(s)
\sum_{ij}\frac{[1-r_{ij}(s)]^2}{r_{ij}(s)}\,ds\right\} \;,
\end{equation}
where $\sum_{ij}$ runs over the three pairs $xy$, $xz$, and $yz$.
The exponent becomes $k_fS(I_{xy}+I_{xz}+I_{yz})/3$ only for constant $k_f$.
The quantities $I_{xy}$, $I_{xz}$, and $I_{yz}$ denote the three
unweighted temperature-ratio integrals.
For example, the dimensionless quantity $I_{xy}$ is given by:
\begin{equation} \label{xy-int}
I_{xy} = \frac{1}{S} \int\limits_0^S
\frac{\bigl[ 1 - r_{xy}(s) \bigr] ^2}{r_{xy}(s)} \, ds \geq 0
\qquad , \qquad r_{xy}(s) = \frac{\varepsilon_y^2(s)}{\< y^2 \>}
\frac{\< x^2 \>}{\varepsilon_x^2(s)} \;.
\end{equation}
From Eq.~(\ref{t-growth}), the average $e$-folding time
$\tau_{\mathrm{ef},\ell}$ for the laboratory-time emittance ratio
$\varepsilon_{\mathrm{tot}}(t_\ell) / \varepsilon_{\mathrm{tot}}(0)$
is calculated to give
\begin{equation} \label{tau3}
\tau_{\mathrm{ef},\ell}^{-1}
=\frac{1}{9\gamma S}\int_0^S\beta_f(s)
\sum_{ij}\frac{[1-r_{ij}(s)]^2}{r_{ij}(s)}\,ds
\simeq\frac{\beta_f}{9\gamma}(I_{xy}+I_{xz}+I_{yz}) \;.
\end{equation}
The last approximation again requires an effectively constant collision rate.
These nonnegative growth rates refer to the rms-emittance product in the
uncoupled scalar model with negligible directional field residuals.
Temperature equality is sufficient to make its collision contribution
vanish, but does not by itself establish full thermodynamic equilibrium.
If a periodic focusing pattern repeatedly restores temperature differences,
the model predicts repeated growth while the collision rate remains finite.
It does not imply a constant exponential rate or rule out all possible
asymptotic saturation mechanisms in an actual ring.

\section{THE EMITTANCE EQUATION FOR COASTING BEAMS IN STORAGE RINGS}
\label{coasting}
For a stationary, azimuthally uniform coasting beam, the longitudinal
phase is uniform and independent of the momentum deviation.
We use $\delta=\Delta p/p$ as the longitudinal kinetic variable.
The relation to revolution-time deviation is
$$
z'=-\frac{\Delta\tau}{\tau}=\eta\delta \;.
$$
An unwrapped longitudinal coordinate is not a bounded stationary random
variable on a ring.
A fixed phase interval may be used to assign a constant $\<z^2\>$ as
bookkeeping for emittance ratios, but the momentum spread itself does
not depend on that interval.
For such a uniform phase convention, $\<z\delta\>=0$ and
$$
\bar\varepsilon_z^2=\frac{\<z^2\>\<\delta^2\>}{\gamma^2},
\qquad
kT_{z;b}=mc^2\beta^2\<\delta^2\> \;.
$$
In particular, the longitudinal kinetic weight is $\gamma^{-2}$,
not $|\eta|$.
The sign of $\eta$ belongs to the orbital phase dynamics rather than
to the positive local random kinetic energy.

The coupled set of equations (\ref{epsderi3}) describing
the emittance changes can now be written as:
\begin{eqnarray}\label{eps3-d}
\frac{1}{\<x^2\>} \frac{d}{ds} \varepsilon_x^2(s) & = &
-\frac{2k_f}{3} \left(
\frac{2\varepsilon_x^2(s)}{\<x^2\>} -
\frac{ \varepsilon_y^2(s)}{\<y^2\>} -
\gamma^{-2}\, \< \delta^2 \> \right) \nonumber \\
\frac{1}{\<y^2\>} \frac{d}{ds} \varepsilon_y^2(s) & = &
-\frac{2k_f}{3} \left(
\frac{2\varepsilon_y^2(s)}{\<y^2\>} -
\frac{ \varepsilon_x^2(s)}{\<x^2\>} -
\gamma^{-2}\, \< \delta^2 \> \right) \\
\gamma^{-2}\, \frac{d}{ds}\< \delta^2 \> & = &
-\frac{2k_f}{3} \left( 2 \gamma^{-2}\, \< \delta^2 \> -
\frac{ \varepsilon_x^2(s)}{\<x^2\>} -
\frac{ \varepsilon_y^2(s)}{\<y^2\>} \right) \; , \nonumber
\end{eqnarray}
Here we neglect the directional field residuals and transverse dispersion,
and retain the scalar, affine-flow closure.
These are local kinetic-moment equations, not full lattice-averaged
intra-beam-scattering formulas.
If we add the Eqs.~(\ref{eps3-d}), all right-hand side terms cancel,
yielding:
\begin{equation} \label{piwi1}
\frac{1}{\<x^2\>} \frac{d}{ds} \varepsilon_x^2(s) +
\frac{1}{\<y^2\>} \frac{d}{ds} \varepsilon_y^2(s) +
\gamma^{-2}\, \frac{d}{ds} \< \delta^2 \> = 0 \;.
\end{equation}
If the transverse rms-sizes are held fixed in addition to the
reference $\gamma$, Eq.~(\ref{piwi1}) can be rewritten as:
$$
\frac{d}{ds} \left[
\frac{\varepsilon_x^2(s)}{\<x^2\>} +
\frac{\varepsilon_y^2(s)}{\<y^2\>} +
\gamma^{-2}\, \< \delta^2 \> \right] = 0 \;.
$$
Integration then gives a constant of this frozen-size kinetic model,
\begin{equation} \label{piwi2}
\frac{\varepsilon_x^2(s)}{\<x^2\>} +
\frac{\varepsilon_y^2(s)}{\<y^2\>} +
\gamma^{-2}\, \< \delta^2 \> = \mathrm{const.}
\end{equation}
This is not Piwinski's storage-ring invariant\cite{piwinski}.
His invariant includes the relation between betatron motion, dispersion,
and momentum deviation, with a longitudinal coefficient whose sign
changes across transition.
The local positive kinetic weight in (\ref{piwi2}) must not be used to
replace that orbital coefficient by its absolute value.

When the transverse sizes vary, (\ref{piwi1}) is still the collisional
moment balance, but derivatives of the inverse sizes prevent it from
being the total derivative of (\ref{piwi2}).
This fact alone neither proves the absence of an equilibrium in a
strong-focusing ring nor invalidates invariants derived under different
orbital assumptions.
For a quantitative ring calculation, the betatron covariance must be
separated from the dispersive part, for example by
$x=x_\beta+\mathcal{D}_x(s)\delta$.
The covariances involving $x$, $x'$, and $\delta$, as well as their
collision-driven changes, then enter the growth rates.

Within the deliberately zero-dispersion reduction, the laboratory
rate of the rms-emittance product is given by (\ref{tau3}), with the
physical temperature ratios defined above.
For a coasting beam this is equivalently the rate of
$(\varepsilon_x\varepsilon_y\sqrt{\<\delta^2\>})^{1/3}$,
since the fixed phase-interval and $\gamma$ factors cancel in a ratio.
A lattice-independent assertion of symmetric heating above and below
transition does not follow from these reduced equations.

\section{RESULTS OF NUMERICAL CALCULATIONS} \label{numeric}
\subsection{Quadrupole Channels} \label{quadchannel}
\begin{figure}[t]
%\vspace*{75mm}
\epsfig{file=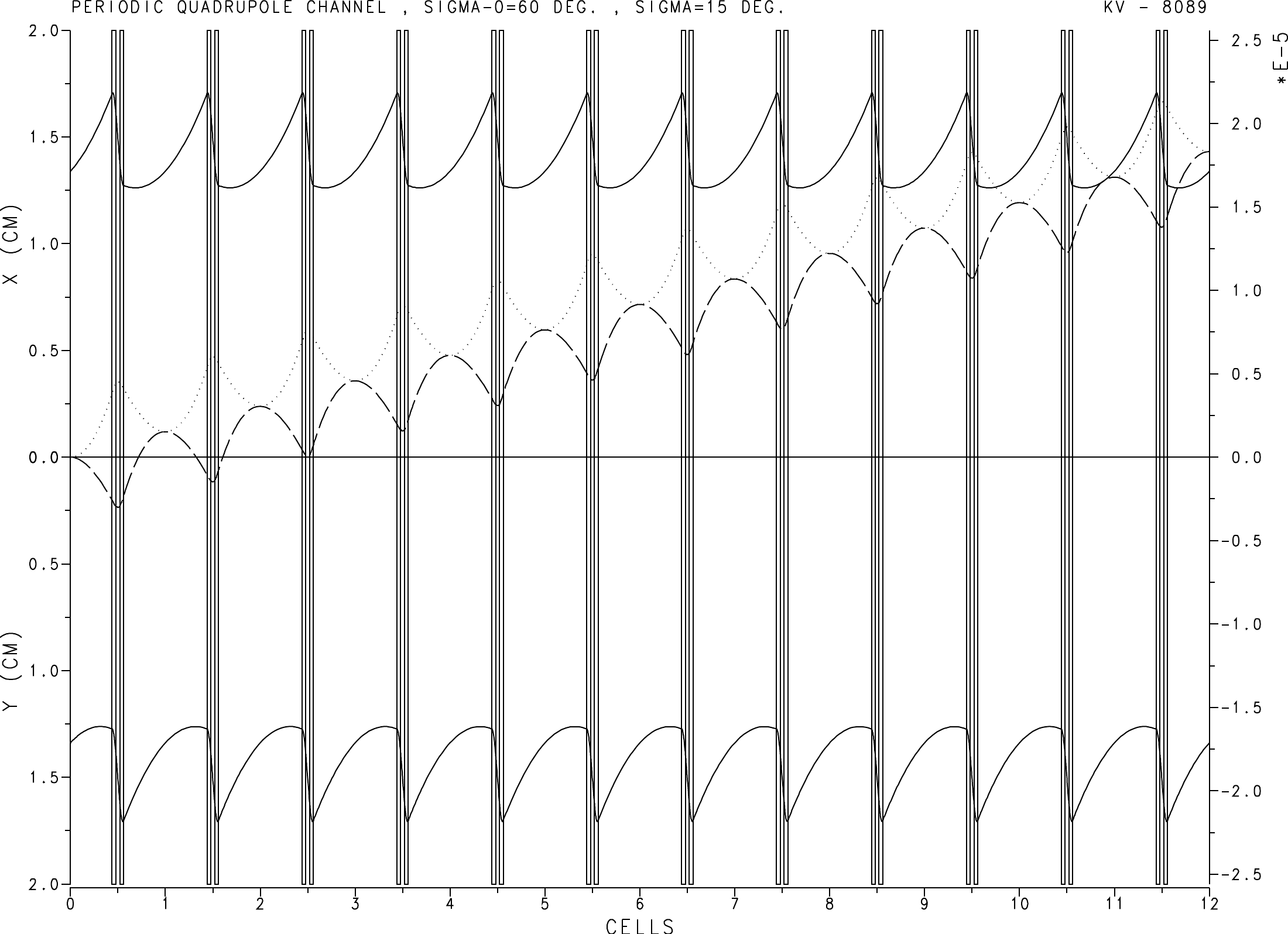,width=125mm,height=75mm}
\caption{Original envelope and emittance growth functions of a
matched beam passing through a Periodic Quadrupole Channel
at $\sigma_0 = 60^\circ$, $\sigma = 15^\circ$.
(The scale on the right-hand side applies to
the dimensionless emittance growth functions.)}
\label{sim-quad}
\end{figure}
With ordinary alternating-gradient quadrupole arrangements, a matched
beam generally does not remain round throughout the focusing period.
Under the additional approximately equal-emittance condition, this
produces a positive $I_e$ in the transverse relaxation model.
It leads to a positive instantaneous emittance-product growth rate
where the collision rate is finite.
The rate can still change as the beam expands or its temperature changes.

The original figures and tables are retained in this section as a record
of the earlier numerical examples.
They are not new integrations of the corrected equations.
The supplied material does not specify the full lattice functions,
density averaging, or numerical data needed for such a recalculation.

According to (\ref{e-fold}), the growth rate
depends on the beam parameters, yielding a specific
value of the `dynamical friction coefficient'
$\beta_f$ (\ref{friction}), and the matched lattice--beam solution, which
determines the `ellipticity' $I_e$.
\begin{table}[htbp]
\begin{center}
\begin{tabular}{|ll|} \hline
ion & $^{40}$Ar$^{1+\vphantom{^X}}$ \\
energy & $4.75$~keV/amu \\
period length $S$ & $4.0$~m \\
zero current phase advance $\sigma_0$ & $60^\circ$ \\
depressed phase advance $\sigma$ & $15^\circ$ \\
beam current $I$ & $0.25$~mA \\
initial rms emittances $\varepsilon_{x,y}(0)$ & $3.125\times 10^{-6}$~m \\
ellipticity $I_e$ (matched beam) & $0.135$ \\
adopted beam-frame coefficient $\beta_f$ & $11.1$~s$^{-1}$ \\
laboratory $e$-folding time $\tau_{\mathrm{ef},\ell}$ & $2.7$~s \\ \hline
\end{tabular}
\caption{Original parameters for the Periodic Quadrupole Channel example.
The growth time is an arithmetic estimate using the retained coefficient
and envelope integral, not a recalculated collision rate.}
\label{quadtab}
\end{center}
\end{table}
An example of a low energy beam transport channel is
plotted in Fig.~\ref{sim-quad}.
The original matched-envelope calculation gives $I_e=0.135$.
Keeping that integral and the listed effective beam-frame rate fixed,
Eq.~(\ref{e-fold}) gives $\tau_{\mathrm{ef},\ell}=2.669$~s.
At $4.75$~keV/amu, $\gamma=1.0000051$, so the frame correction is negligible
at the precision of the original $2.7$~s estimate.
This time corresponds to approximately $6.39\times10^5$ focusing periods.

The growth rate of the transverse emittances
per period amounts to
$$
\frac{\varepsilon_\perp(S)}{\varepsilon_\perp(0)} - 1
\approx 1.565\times10^{-6} \;.
$$
\subsection{Storage Rings}
\begin{figure}[t]
%\vspace*{75mm}
\epsfig{file=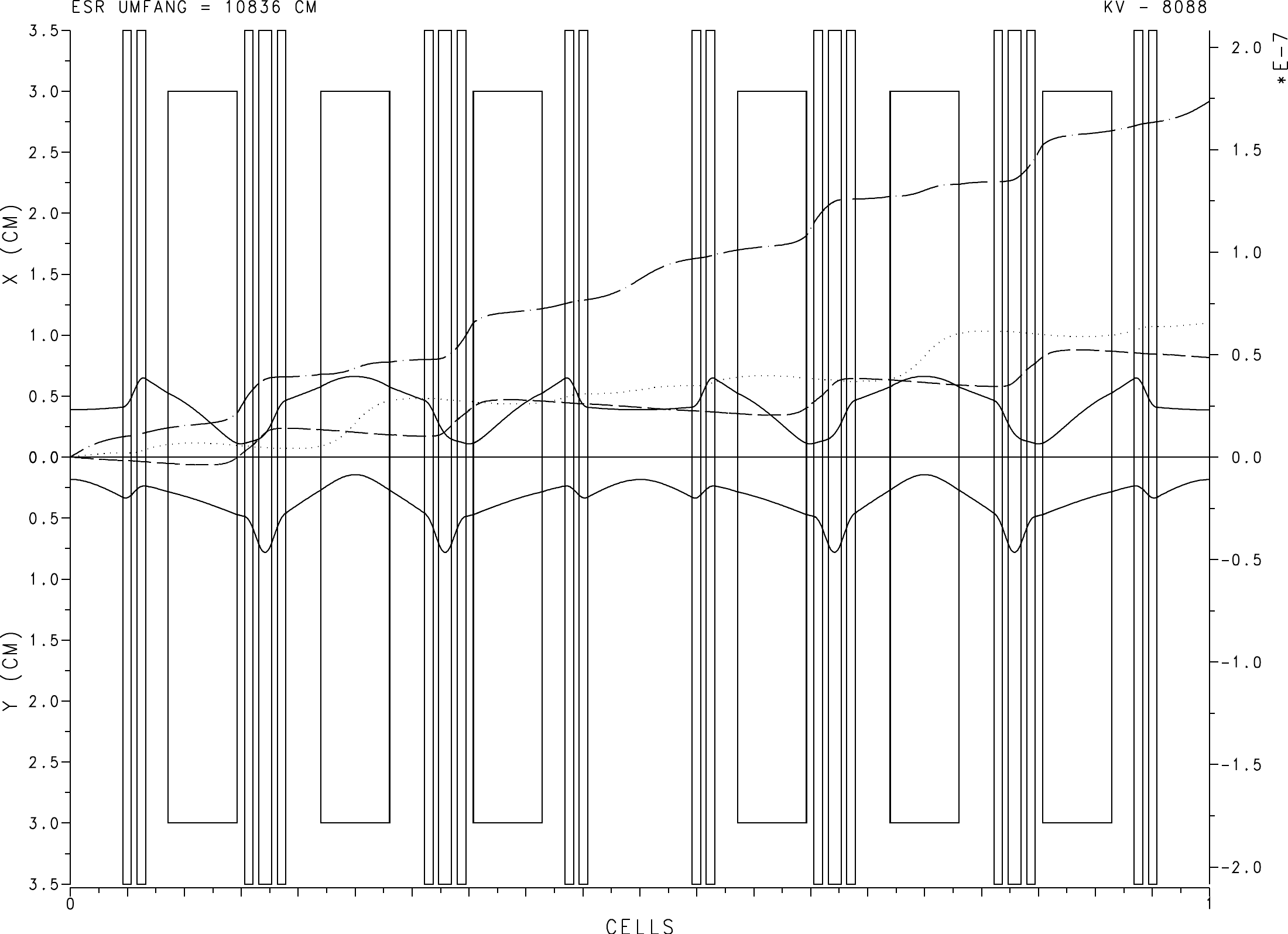,width=125mm,height=75mm}
\caption{Original envelope and emittance growth functions of a
beam passing through the `Experimental Storage Ring' (ESR)
at $Q_h = 2.31$ and $Q_v = 2.25$.
(The scale on the right-hand side applies to
the dimensionless emittance growth functions.)}
\label{sim-esr}
\end{figure}
The emittance growth effect due to temperature imbalance is not
negligible, if the time the beam resides within the focusing structure
is of the order of the $e$-folding time (\ref{e-fold}) or more.
This is the case in circular devices such as storage rings.

We base our calculations on the geometry and the beam data
of the `Experimental Storage Ring' (ESR)\cite{franzke,franzetal},
which started operation at GSI in 1990.

The original calculation neglected the dispersive contribution to the
transverse beam widths.
In a refined ring model the local dispersion $\mathcal{D}_x(s)$ and the
associated momentum correlations are required, not only a modified envelope.
The replacement of the old longitudinal temperature by the physical
kinetic temperature also changes the $xz$ and $yz$ temperature-ratio integrals.
Thus the original numerical rates cannot be promoted to corrected ring
predictions without recomputing the collision coefficients and the lattice averages.

\begin{table}[htbp]
\begin{center}
\begin{tabular}{|ll|} \hline
\mbox{ion} & $^{20}$Ne$^{10+\vphantom{^X}}$ \\
energy & $250$~MeV/amu \\
period length $S$ & 108.36 m \\
horizontal tune $Q_h$ & $2.31$ \\
vertical tune $Q_v$ & $2.25$ \\
$\eta = \gamma^{-2} - \gamma_t^{-2}$ & $0.493$ \\
beam current $I$ & $7$~mA \\
initial rms emittances $\varepsilon_{x,y}(0)$ & $0.25\times 10^{-6}$~m \\
initial rms momentum spread $\Delta p/p$ & $1.3\times 10^{-4}$ \\
original ellipticity $I_{xy}$ & $4.710$ \\
original ellipticity $I_{xz}$ & $2.689$ \\
original ellipticity $I_{yz}$ & $2.382$ \\
original quoted coefficient $\beta_f$ & $0.1306$~s$^{-1}$ \\
reported horizontal emittance time & $10.5$~s \\
reported vertical emittance time & $13.7$~s \\
reported momentum-variance time & $ 1.9$~s \\
reported total emittance time   & $ 7.0$~s \\ \hline
\end{tabular}
\caption{Original ESR parameters and reported numerical outputs, retained for comparison.
The $1.9$~s entry refers to $\<\delta^2\>$ rather than its square root.
These growth times are not recalculated predictions of Eqs.~(\ref{tau3}) and (\ref{eps3-d}).}
\label{esrtab}
\end{center}
\end{table}

The original beam parameters correspond to approximately
$2.6\times10^9$ stored particles.
For $250$~MeV/amu, the reference factors are
$\gamma\simeq1.26839$ and $\beta\simeq0.61516$.
The laboratory revolution time is therefore
$$
T_{\mathrm{rev}}=\frac{108.36~\mathrm m}{c\beta}
\simeq5.876\times10^{-7}~\mathrm s \;.
$$
The original calculation reported
$$
\frac{\varepsilon_x(S)}{\varepsilon_x(0)}-1\simeq5.6\times10^{-8},\qquad
\frac{\varepsilon_y(S)}{\varepsilon_y(0)}-1\simeq4.3\times10^{-8},
$$
$$
\frac{\<\delta^2\>(S)}{\<\delta^2\>(0)}-1\simeq3.0\times10^{-7} \;.
$$
These increments give laboratory growth times of approximately
$10.5$~s and $13.7$~s for the transverse emittances and
$1.96$~s for the momentum variance.
Since $\bar\varepsilon_z\propto\sqrt{\<\delta^2\>}$ in the coasting-beam
convention, its inferred growth time is instead $3.92$~s.
Combining the increments with this square-root relation gives
approximately $7.08$~s for the total emittance, consistent with the
reported rounded value of $7.0$~s and about $1.2\times10^7$ turns.
This is a consistency check of the reported increments, not a validation
of the physical model that produced them.

There is also a separate time-frame issue in the original coefficients.
If $0.1306~\mathrm{s}^{-1}$ is treated as a beam-frame rate and the old
integrals are frozen, their sum $9.781$ gives a beam-frame time of
$7.046$~s and a laboratory time of $8.94$~s.
The latter is only a frame-conversion illustration.
It is not a corrected ESR prediction, since the old integrals and
coefficient have not been recomputed using the new kinetic definitions.
The sizable original ellipticities also do not establish that the
near-isotropic scalar closure is quantitatively controlled for this example.

Within the reduced model, reducing the collision-rate-weighted temperature
imbalances reduces the instantaneous emittance-product growth rate.
Optimizing a real ring additionally requires the full dispersive optics
and the dependence of the collision rate on beam density and temperature.
Smooth transverse envelopes alone do not determine the optimum.

\subsection{Computer Simulations of Charged Particle Beams}
\begin{figure}[t]
%\vspace*{70mm}
\epsfig{file=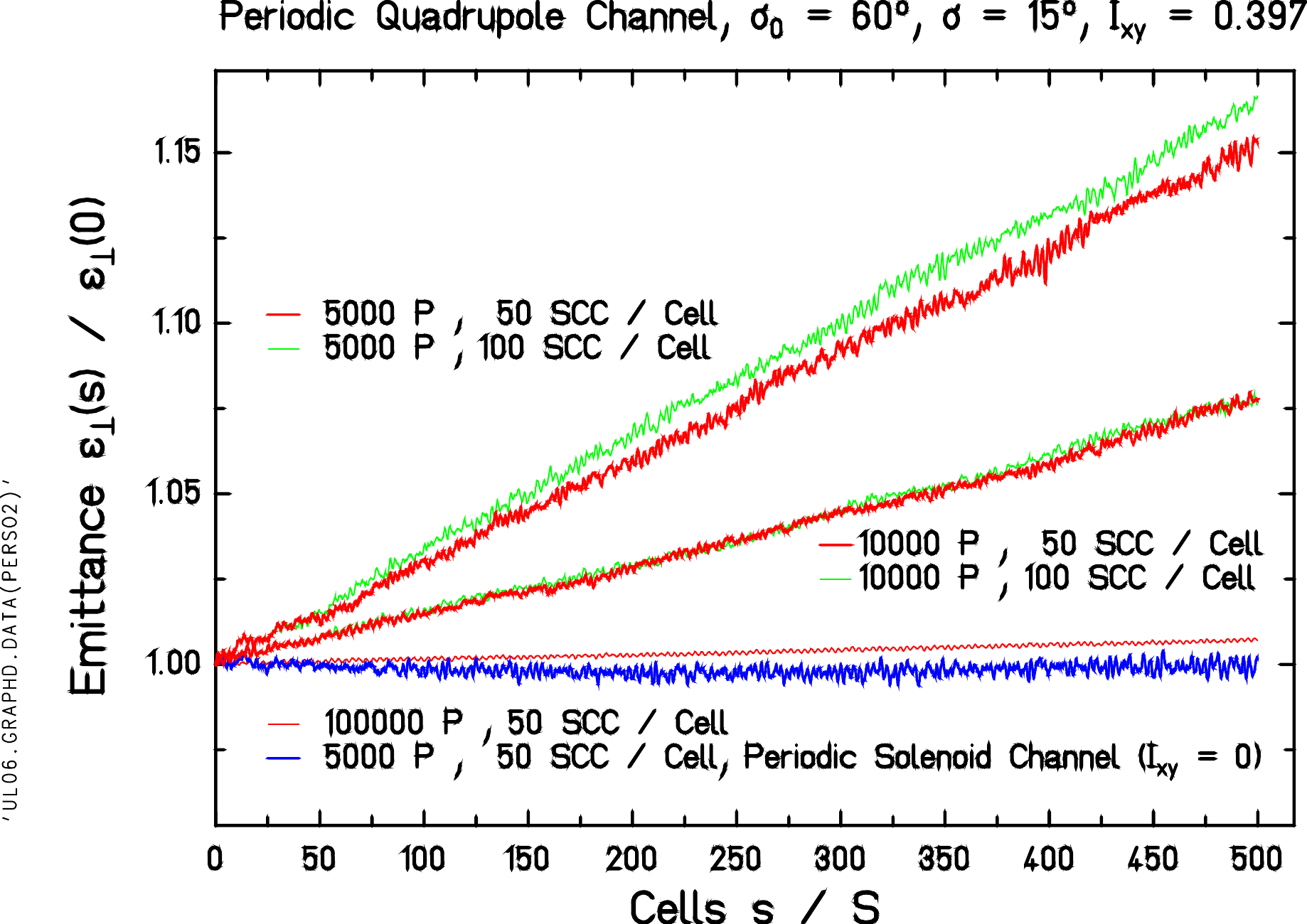,width=125mm,height=70mm}
\caption{Emittance growth functions obtained by `particle-in-cell'
simulations of matched beams passing through different types of
beam transport channels at $\sigma_0 = 60^\circ$, $\sigma = 15^\circ$.
The original curves are retained.
The plot is labeled $I_{xy}=0.397$, whereas Tab.~\protect{\ref{quadtab}}
lists $I_e=0.135$.}
\label{sim-old}
\end{figure}
The Fokker-Planck approach to investigate the behavior
of charged particle beams improves our capability not only
to analyze effects observed in reality, but also to interpret
the results of computer simulations of such beams.

The joint effect of all simplifications necessary to keep
the computing time finite can be visualized
in the way that an additional `error field' is added to
the `true field' we would get in the ideal case.
In other words, the calculated self-field $\vec{E}_c$
can be interpreted as the sum of the true self-field
$\vec{E}_t$ generated by the real beam and a
statistically fluctuating `error field' $\vec{E}_f$:
$$
\vec{E}_c = \vec{E}_t + \vec{E}_f \;.
$$
If the fluctuating error field is short-correlated and produces small
momentum increments, a drift--diffusion model may approximate its action.
The general moment equations (\ref{deri}) then remain applicable.
Numerical errors, however, can also contain systematic or correlated components.
They do not automatically produce the same drift as elastic self-collisions,
and their diffusion need not satisfy (\ref{diffu}).

The original discussion quoted a representative fractional growth per cell
of approximately $4\times10^{-5}$ and compared it with the earlier
physical estimate $1.5\times10^{-6}$.
Their arithmetic ratio is
$$
\frac{1.5\times10^{-6}}{4\times10^{-5}}=0.0375\simeq0.04 \;.
$$
It is a friction-rate ratio only if both cases share the same
relaxation closure and collision-weighted temperature-ratio function.
Those conditions are not established by the retained plot.
The plot contains several particle numbers and space-charge step counts,
and the quoted rate is not tied to a specified trace or fitting interval.
Its ellipticity label also differs from the value in Tab.~\ref{quadtab}.
A quantitative calibration of $\beta_f^{\mathrm{sim}}$ therefore requires
the original simulation data or a new convergence study.

Figure~\ref{sim-old} illustrates that numerical emittance growth can depend
strongly on particle number and focusing configuration.
The nearly stationary solenoid trace is a result of that numerical example.
It is not a general consequence of $I_e=0$, since additive numerical
diffusion can increase emittance even when the transverse temperatures are equal.
A round, uncoupled beam with $I_e=0$ has no temperature-balancing growth
only within the self-collision relaxation model.
Solenoidal transport with angular momentum or transverse coupling requires
an appropriate Larmor-frame or full-covariance treatment.

\section{CONCLUSIONS}
The collisionless Vlasov equation describes Hamiltonian mean-field
transport of the one-particle distribution.
A Fokker-Planck operator adds a reduced description of short-correlated
small momentum changes when a diffusion approximation is justified.
This distinction does not imply a violation of Liouville's theorem for
the full Hamiltonian many-particle system.

Applying Sacherer's\cite{sacherer} moment method to the kinetic equation
gives additional drift and diffusion contributions to the beam moments.
A prescribed bath drag produces an explicit envelope-damping term.
For momentum- and energy-conserving self-collisions, the drag is instead
relative to the local mean flow and there is no direct damping force on that flow.
With an affine, uncoupled mean flow and a spatially uniform effective
relaxation rate, the emittance equations describe temperature balancing.
Their scalar rate is a near-equilibrium approximation to Coulomb transport.
Its density and approximate charge-state dependence follow from the
chosen scattering model, with the temperature and Coulomb logarithm specified.

For the restricted transverse model, negligible directional field residuals,
and approximately equal emittances, a constant beam-frame coefficient gives
$$
\tau_{\mathrm{ef},\ell}=\frac{4\gamma}{\beta_f I_e} \;.
$$
The `ellipticity' $I_e$ describes the matched lattice--beam solution,
not generally the lattice alone.
Persistent temperature imbalance gives positive instantaneous growth of
the emittance product in this model.
It does not imply a parameter-independent, indefinitely exponential growth law.
When the coefficient varies along the period, the weighted integral in
Eq.~(\ref{e-fold}) must be used.

The three-dimensional extension requires the physical longitudinal
random momentum.
For constant nonzero $\eta$, the appropriate rescaled emittance is
$\bar\varepsilon_z=\varepsilon_z/(\gamma|\eta|)$.
For coasting beams its kinetic variance is proportional to
$\<\delta^2\>/\gamma^2$, not $|\eta|\<\delta^2\>$.
The scalar three-temperature closure gives
$$
\tau_{\mathrm{ef},\ell}=
\frac{9\gamma}{\beta_f(I_{xy}+I_{xz}+I_{yz})}
$$
when the coefficient is effectively constant over the period.
The weighted version is Eq.~(\ref{tau3}).
These are rms-product results within an uncoupled closure, rather than
a general entropy theorem or a complete storage-ring scattering theory.

The field-energy discussion must retain the force moments of the
uniform reference distribution separately from those of the actual beam.
A constant total excess energy does not justify discarding every
directional field residual.
Likewise, a quantitative storage-ring calculation must retain dispersion
and the appropriate covariance dynamics.
The sign of phase slip remains important above and below transition.

The original numerical examples have been retained with their historical status
made explicit, the momentum-variance growth time correctly identified,
and the time-frame convention distinguished from the reported outputs.
They require recalculation before being used as quantitative predictions
of the corrected kinetic model.
Numerical particle noise may admit a Fokker-Planck description, but its
coefficients and fluctuation--dissipation properties require independent calibration.
\acknowledge
I wish to thank I.~Hofmann for frequent valuable
discussions during the course of this work.
I am also grateful to K.~Johnson who -- during
his leave from Los Alamos at GSI -- read the
manuscript and offered helpful criticism.

\end{document}